\documentclass[
    ,final            
  ,numberedheadings 
  ]
  {aipproc}

\layoutstyle{8x11single}

\begin{document}

\title{A Correlation of Spectral Lag Evolution with Prompt Optical Emission in GRBs?\footnote{Adapted from a contribution to the Proceedings of the 2008 Nanjing GRB Conference. Edited by Y. F. Huang, Z. G. Dai and B. Zhang.}}

\classification{95.75.Wx, 95.85.Kr, 95.85.Pw and 98.70.Rz}
\keywords      {gamma rays: bursts, radiation mechanisms: non-thermal, temporal analysis.}

\author{Michael Stamatikos}{
  address={Astroparticle Physics Laboratory, Code 661, NASA/Goddard Space Flight Center, Greenbelt, MD 20771 USA\footnote{Correspondence to Michael.Stamatikos-1@nasa.gov.}}
}

\author{Tilan N. Ukwatta}{
  address={The George Washington University, Washington, DC 20052 USA}
}

\author{Taka Sakamoto}{
  address={CRESST, University of Maryland, Baltimore County, Baltimore, MD 21250 USA}
}

\author{Kalvir S. Dhuga}{
  address={The George Washington University, Washington, DC 20052 USA}
}

\begin{abstract}
We report on observations of correlated behavior between the prompt $\gamma$-ray and optical emission from GRB 080319B, which (i) strongly suggest that they occurred within the same astrophysical source region and (ii) indicate that their respective radiation mechanisms were most likely dynamically coupled. Our preliminary results, based upon a new cross-correlation function (CCF) methodology for determining the \emph{time-resolved} spectral lag, are summarized as follows. First, the evolution in 
the arrival offset of prompt $\gamma$-ray photon counts between Swift-BAT 15-25 keV and 50-100 keV energy bands \emph{(intrinsic $\gamma$-ray spectral lag)} appears to be anti-correlated with the arrival offset between prompt 15-350 keV $\gamma$-rays and the optical emission observed by TORTORA \emph{(extrinsic optical/$\gamma$-ray spectral lag)}, thus effectively partitioning the burst into two main episodes at $\sim T+28\pm2$ sec. Second, prompt optical emission is nested within intervals of (a) trivial intrinsic $\gamma$-ray spectral lag ($\sim T+12\pm2$ and $\sim T+50\pm2$ sec) with (b) discontinuities in the hard to soft evolution of the photon index for a power law fit to 15-150 keV Swift-BAT data ($\sim T+8\pm2$ and $\sim T+48\pm1$ sec), both of which coincide with the rise ($\sim T+10\pm1$ sec) and decline ($\sim T+50\pm1$ sec) of prompt optical emission. This potential discovery, robust across heuristic permutations of BAT energy channels and varying temporal bin resolution, provides the first observational evidence for an implicit connection between spectral lag and the dynamics of shocks in the context of canonical fireball phenomenology.
\end{abstract}

\maketitle


\section{Introduction}

Swift's unique dynamic response and spatial localization precision, in conjunction with correlative ground-based follow-up efforts, has resulted in the collaborative broad-band observations of GRB 080319B \cite{Racusin:2008c}. In the context of an analysis focused on confronting the lag-luminosity relation \cite{Norris:2000b} in the Swift era \cite{Stamatikos:2008b}, a correlation was observed between the evolution of time-resolved spectral lag and the behavior of the extraordinarily well-sampled prompt optical emission light curve associated with GRB 080319B. In general, the spectral lag is determined via either a peak pulse fit \cite{Norris:2005b} or cross-correlation function (CCF) analysis \cite{Band:1997b}. Previous studies have reported on the variability of spectral lag throughout burst emission \cite{Chen:2005}, as well as its correlation to pulse evolution \cite{Hakkila:2008}. In this work, we develop a new method to calculate the time-resolved spectral lag via a modification to the traditional CCF approach. In this manner, we are able to explore for the first time the evolution and correlation of the intrinsic $\gamma$-ray spectral lag with prompt optical emission. We interpret these correlated behaviors as strong observational evidence that the prompt optical and $\gamma$-ray emission of GRB 080319B took place within the same astrophysical source region, with indications that their respective radiation mechanisms where dynamically coupled throughout the prompt phase of the burst.

\section{Methodology}

The Burst Advocate $\texttt{(BA)}$ script was used to generate BAT event data and quality maps, which were used to construct 32 ms binned\footnote{ A binning of 32 ms was used based upon a heuristic exploration of various bin resolutions (e.g. 4 ms, 8 ms, 16 ms, 32 ms, 64 ms, etc.). Once a spectral lag measurement was made, it was reproduced using light curves of several adjacent bin resolutions. In principle, the binning of the light curve should be smaller than the spectral lag in order to resolve the CCF peak.} (mask-weighted\footnote{Although one increases the signal-to-noise ratio (SNR) using non-mask-tagged (raw, non-background subtracted) light curves, artifacts either intrinsic to the detector or associated with the background may mimic signal pulses and thus be treated as such in a CCF analysis. Also, the background may not be trivial to model since it may be highly variable due to slewing. Hence, in order to be conservative and consistent with future studies, background subtracted, i.e. mask-tagged, light curves were used throughout the analysis.}, background-subtracted) light curves (temporal photon spectra) for $E_{A}=15-25$ keV and $E_{B}=50-100$ keV energy bands\footnote{Permutations of canonical BAT channel (1-4) light curve pairings were investigated. Ultimately BAT channels 1 (15-25 keV) and 3 (50-100 keV) were used in the analysis since they represented the largest energy differential with the highest count rate significance.}, via the $\texttt{BATBINEVT}$ analysis task. The resultant $\texttt{FITs}$ files were analyzed via the CCF method \citep{Band:1997b} in order to quantify the temporal correlation between the two series of GRB light curves in differing energy channels, i.e. $E_{A}$ and $E_{B}$, via the following:

\begin{equation}
CCF\left(t_{o},\nu_{E_{A}},\nu_{E_{B}}\right)=\frac{\langle\nu_{E_{A}}(t)\nu_{E_{B}}(t+t_{o})\rangle}{\sigma_{\nu_{E_{A}}}\sigma_{\nu_{E_{B}}}}\equiv CCF_{AB},
\label{CCF}
\end{equation}
where $\sigma_{\nu_{E_{n}}}=\langle\nu_{E_{n}}^{2}\rangle^{1/2}$. The CCF between $E_{A}$ and $E_{B}$ peaks at a given temporal offset $\left(t_{o}\right)$, known as the \emph{spectral lag} $\left(\tau_{AB}\right)$, which is defined as positive if the systematic shift in the arrival times of photon counts between pairs of light curves results in higher energy $\left(E_{B}\right)$ photons arriving \emph{before} those of lower energy $\left(E_{A}\right)$. The generic CCF, based upon the Pearson Correlation, is defined in Equation~\ref{CCF}. If the mean is subtracted\footnote{Mean subtraction is the formalism identical to that used by the \texttt{C\_CORRELATE} function of \texttt{IDL 6.2}. The formal definition may be found online (\texttt{http://c1.dmf.arm.gov/base/idl\_6.2/C\_CORRELATE.html}).}, then $v(t) = d(t) -\langle d(t)\rangle$. If the mean is not subtracted, then $v(t) = d(t) - b(t)$, where an attempt is made to remove background counts $b(t)$. Our tests indicated that both methods agreed when one calculates the \emph{time-averaged} spectral lag, i.e. the spectral lag determined over the entire duration of the GRB. This included a test of subtracting the mean of a light curve that was nested within two large background intervals, which had the effect of reducing the mean and increasing the correlation amplitude. Note that if one takes an infinite interval, the mean would go to background, which fluctuates about zero, resulting in effectively not subtracting the mean in the first place, as prescribed for transient sources, i.e. GRBs \cite{Band:1997b}.

However, when one tries to extract a \emph{time-resolved} spectral lag, i.e. the spectral lag over a segment of the light curve, then only the mean subtraction method consistently works. This is a consequence of edge effects introduced by the assumptions of a stationary versus a transient temporal signal characterization. If one subtracts the mean, there is an assumption that the signal exists as a stationary wave throughout time, i.e. beyond the sampled time series. If one does not subtract the mean, then the assumption is that the signal is entirely contained within the time series. Both methods disregard data beyond the overlapped interval due to the introduction of a spectral lag shift.

Hence, for a bright GRB like 080319B, a non-mean subtracted time-resolved spectral lag analysis is dominated by the segment of the light curve described by a square wave, and the real light curve signal is treated as fluctuations. A mean-subtracted time-resolved spectral lag analysis will extract the spectral lag since it smoothes out the amplitude and focuses on the signal variance. Hence, we utilized the mean-subtracted definition of the CCF, since it is both more conservative and robust. Although it has been demonstrated that spectral lag variability is ubiquitous in GRBs \cite{Chen:2005,Hakkila:2008}, previous methods required relatively well-behaved (FRED-like) sub pulses. Our new methodology treats structure without regard to its functional form and affords greater flexibility for time-resolved spectral lag evolution studies.

\begin{figure}
\includegraphics[height=.54\textheight]{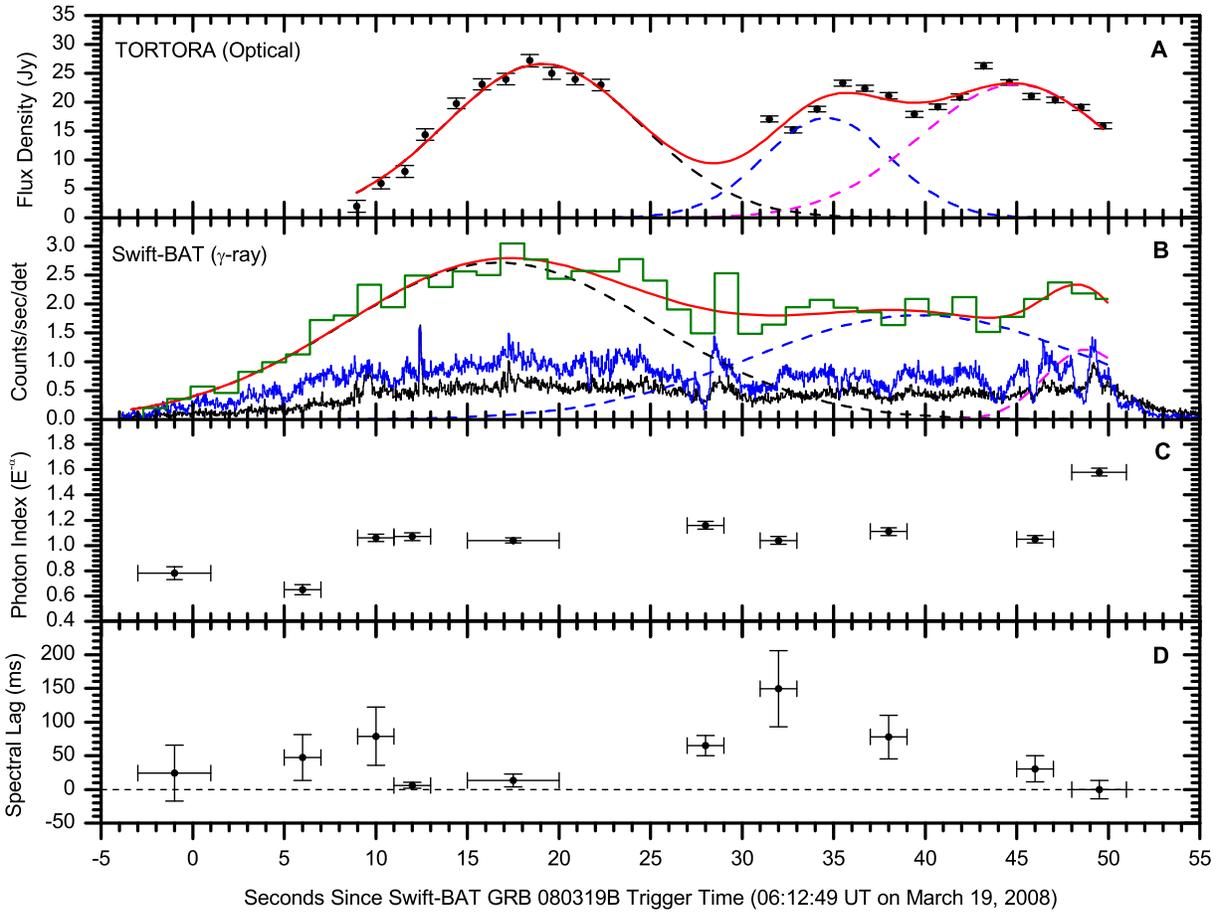}
\caption{\emph{\textbf{Panel A -}} Prompt optical flux density as observed by TORTORA for $\sim$1.3 second exposure intervals (black circles). A cumulative fit (solid red line) is based upon 3 Gaussian dashed curves with peaks at $\sim$T+19 sec (black), $\sim$T+34 sec (blue) and $\sim$T+45 sec (magenta). \emph{Data has been digitized from Figure 1 of \cite{Racusin:2008c}}. \emph{\textbf{Panel B -}} Swift-BAT prompt $\gamma$-ray (32 ms bin) light curves for canonical energy channels 1 (15-25 keV, solid black) and 3 (50-100 keV, solid blue). 15-350 keV summed, 1.3 second bin light curve illustrated in solid green. A cumulative fit (solid red line) is based upon 3 Gaussian dashed curves with peaks at $\sim$T+17 sec (black), $\sim$T+40 sec (blue) and $\sim$T+49 sec (magenta). \emph{\textbf{Panel C -}} Photon index $(E^{-\alpha})$ for power law fit to 15-150 keV Swift-BAT data, with $1\sigma$ error bars. Note that hard to soft evolution occurs in discontinuous steps at $\sim T+8\pm2$ and $\sim T+48\pm1$ sec, which coincide with the respective rise ($\sim T+10\pm1$ sec) and decline ($\sim T+50\pm1$ sec) of prompt optical emission \emph{(see Panel A)}. \emph{\textbf{Panel D -}} Time-resolved, intrinsic spectral lag between BAT 32 ms bin light curves for 15-25 keV and 50-100 keV energy bands, based upon a Gaussian fit to the CCF peak, with $1\sigma$ error bars derived from 1000 Monte Carlo bootstrapped simulations. Intervals of trivial spectral lag occur at $\sim T+12\pm2$ and $\sim T+50\pm2$ sec and also coincide with the rise and decline of optical emission. Overall, the temporal evolution of the time-resolved spectral lag can be described by two main episodes, where the intrinsic (BAT Channel 1/Channel 3) $\gamma$-ray spectral lag is greater for t > $\sim T+28\pm2$ sec. Note that this is anti-correlated with the extrinsic optical/$\gamma$-ray spectral lag observed via the smoothed Gaussian fits to the optical emission, where the spectral lag is of the order of a few seconds at t < $\sim T+28\pm2$ seconds, and either zero or negative at later times (ambiguity due to peak misalignment). Such correlated spectral and temporal behavior indicates that the prompt optical and $\gamma$-ray emission took place within the same source region, with radiation mechanisms that were most likely dynamically coupled.}
\label{plot}
\end{figure}

In order to test the viability of an interval, a large known spectral lag was artificially introduced, which shifted one of the test light curve pairs. Intervals where the the analysis was unable to recover the artificial spectral lag due to either low SNR or lack of structure were not used. The CCF method, regardless of definition, is sensitive to the SNR and, more importantly, to the shape of the region. Morphological tests using simple square, triangular and Gaussian pulse shapes have illustrated that both mean and non-mean subtracted CCF definitions failed to recover the artificial spectral lag that was obscured in regions where (i) only slopes were compared, (ii) SNR was low and/or (iii) error bars were large. This puts a fundamental limit on determining the time-resolved spectral lag for discrete intervals, which precluded an arbitrary systematic sampling of GRB 080319B, as illustrated in Figure~\ref{plot}, Panel D. Hence, although our methodology is robust, it does require a selection on intervals with structured variability. This is not an issue for time-averaged results, since the burst's structured variability interval is nested between periods of quiescence.

The error bars on the CCF function were determined via a primary Monte Carlo (MC) simulation, which were constructed from 1000 realizations of light curve pairs, within the $1\sigma$ error bars of the data. The distribution of CCF values within a given bin defined its error bar. Once the error bars were fixed in this manner, a secondary MC simulation constructed a subsequent set of 1000 realizations of light curve pairs, within the $1\sigma$ error bars of the data. Each resulting CCF (with errors fixed from the primary MC simulation) was subjected to a Gaussian fit:

\begin{equation}
y=Ae^{-\left(x-x_{c}\right)^{2}/\left(2w^{2}\right)},
\label{Gaussian}
\end{equation}
whose peak was extracted via the \texttt{IDL} function \texttt{MPFITPEAK\footnote{See \texttt{http://astro.berkeley.edu/$\sim$heiles/idlprocs/mpfit/mpfitpeak.pro}, for more information.}}. The peak fit to the data was used to determine the spectral lag value, while the numerical standard deviation of the distribution of simulated peaks was used to determine the error. We interpret this bootstrapped result as the $1\sigma$ spectral lag error, as illustrated in Figure~\ref{plot}, Panel D.

In general, spectral lag calculations are sensitive to a series of selection parameters that include the energy band pass of each comparative light curve, temporal bin resolution and emission interval. The CCF and pulse fit methods are both based upon a functional expression for the peak fit (Gaussian, cubic, etc.) while the former requires a bin shift range. A self-consistent comparison, which includes uncertainties, must account for these selection effects. Efforts to standardize the spectral lag analysis are ongoing and would facilitate direct comparative studies across detector sensitivities and energy band passes.

\section{Results \& Discussion}

The generic agreement of the overall temporal coincidence and morphology between the prompt $\gamma$-ray and optical light curves (Figure~\ref{plot}, Panels A \& B) suggests that they arose from a common source region. However, separate radiation mechanisms were most likely responsible since the extrapolated $\gamma$-ray flux density to the optical band was deficient by $\sim$4 orders of magnitude when compared to observation \cite{Racusin:2008c,Kumar:2008,Yu:2008}. The steep rise/decline, short duration and lack of increasing pulse width of the prompt optical emission disfavor external forward/reverse shocks. Hence, internal shocks have been suggested as the source region with synchrotron emission responsible for the optical and inverse Compton scattering/synchrotron self Compton for the $\gamma$-rays, with associated GeV photon emission \cite{Racusin:2008c,Kumar:2008}. Alternatively, it has been suggested that non-relativistic forward internal shocks generated the prompt optical emission, while relativistic reverse internal shocks generated the prompt $\gamma$-rays, with sub-GeV/MeV photon emission \cite{Yu:2008}. Such high energy emission may be tested by Fermi (formally known as GLAST) via joint analyses with Swift-BAT \citep{Stamatikos:2008c}.

Our preliminary results are illustrated in Figure~\ref{plot}, Panels A-D. There are several observations that lead to effectively separating the burst's duration into two main episodes partitioned roughly at the midpoint of $\sim T+28\pm2$ sec. The first is that the bimodal evolution of the intrinsic $\gamma$-ray spectral lag increases at t $>\sim T+28\pm2$ sec, which appears to be anti-correlated with the extrinsic (optical/$\gamma$-ray) spectral lag observed via the smoothed Gaussian fits, where the spectral lag is of the order of a few seconds when t < $\sim T+28\pm2$ sec (Panels A \& B). Beyond this common midpoint, the optical and $\gamma$-rays do not correlate as well or at least are ambiguously correlated, i.e. extrinsic (optical/$\gamma$-ray) spectral lag is either zero or negative at later times (ambiguity due to peak misalignment). Hence, the intrinsic time-resolved $\gamma$-ray spectral lag is maximum at t $>\sim T+28\pm2$ sec, while the extrinsic time-resolved (optical/$\gamma$-ray) spectral lag is maximum at t $<\sim T+28\pm2$ sec. In addition, an independent analysis \cite{Margutti:2008} of BAT 15-150 keV light curves has revealed that the characteristic variability timescale of GRB 080319B was $\sim$100 ms for t < $\sim T$+28 sec and $\sim$1 sec for t > $\sim T$+28 sec. Furthermore, time-resolved spectral analysis by Konus-Wind illustrated that E$_{peak}$ decreased from $751\pm26$ keV to $537\pm28$ keV at $\sim T+24\pm2$ sec \cite{Racusin:2008c}.

The hard to soft evolution of the photon index for time-resolved power law fits to 15-150 keV Swift-BAT data occurs in steps at $\sim T+8\pm2$ and $\sim T+48\pm1$ sec, which coincide with the respective rise ($\sim T+10\pm1$ sec) and decline ($\sim T+50\pm1$ sec) of prompt optical emission (see Panels A \& C). This is consistent with Konus-Wind time-resolved spectral analysis and hardness ratios \cite{Racusin:2008c}. We also note that intervals of trivial intrinsic spectral lag coincide with the rise and decline of prompt optical emission (see Panels A \& D), at $\sim T+12\pm2$ and $\sim T+50\pm2$ sec, respectively. We interpret these correlated behaviors as strong observational evidence that the prompt optical and $\gamma$-ray emission took place within the same astrophysical source region, which until now has only been conjecture. We also find indications for a dynamical coupling between the radiation mechanisms, perhaps via the processes mentioned above \cite{Racusin:2008c,Kumar:2008,Yu:2008}.

This potential discovery, provides the first observational evidence for an implicit connection between spectral lag and the dynamics of internal shocks in the context of canonical fireball phenomenology. A full theoretical analysis of this result is currently in preparation. Future work includes an application of our methodology to observations of a subset of bursts with prompt optical emission to probe either the unique or ubiquitous nature of these observations. Ultimately, understanding the mechanism(s) responsible for spectral lag may reveal a fundamental and unprecedented view from within the GRB fireball.



\begin{theacknowledgments}
 The authors are grateful to Craig Markwardt and David Band for very fruitful discussions in regards to this analysis. M. Stamatikos is supported by an NPP Fellowship at NASA-GSFC administered by ORAU.
\end{theacknowledgments}



\bibliographystyle{aipproc}   

\bibliography{Stamatikos_GRB_Nanjing_astroph}

\IfFileExists{\jobname.bbl}{}
 {\typeout{}
  \typeout{******************************************}
  \typeout{** Please run "bibtex \jobname" to optain}
  \typeout{** the bibliography and then re-run LaTeX}
  \typeout{** twice to fix the references!}
  \typeout{******************************************}
  \typeout{}
 }

\end{document}